\documentclass[3p,times,procedia]{elsarticle}
\usepackage{nupha_ecrc}

\volume{00}

\firstpage{1}

\journalname{Nuclear Physics A}

\runauth{R. D. Pisarski, F. Rennecke, S. Valgushev, and A. Tsvelik}

\jid{nupha}

\jnltitlelogo{Nuclear Physics A}

\usepackage{amssymb}

\usepackage[figuresright]{rotating}

\begin{document}

\begin{frontmatter}



\dochead{XXVIIIth International Conference on Ultrarelativistic Nucleus-Nucleus Collisions\\ (Quark Matter 2019)}

\title{The Lifshitz Regime and its Experimental Signals}
\pdfoutput=1

\author[nt]{Robert D. Pisarski}
\address[nt]{Department of Physics, Brookhaven National Laboratory, Upton, NY 11973}
\author[nt]{Fabian Rennecke}
\author[cm]{Alexei Tsvelik}
\address[cm]{Condensed Matter Physics and Material Sciences Division, Brookhaven National Laboratory, Upton, NY 11973}
\author[nt]{Semeon Valgushev}

\begin{abstract}
  We discuss the possibility of a Lifshitz regime, where the dispersion relation for Goldstone bosons and related
  fields has a minimum at nonzero momenta.  Studies with the 
  Functional Renormalization Group suggest that this occurs over a wide region in the plane of temperature and
  baryon chemical potential.  Conversely,
  the FRG finds that the region in which fluctuations from a critical endpoint are significant
  is rather small.  We suggest that this is due generically to the narrowness of the tricritical region in the chiral limit.
  Even if particles are produced in thermal equilibrium, a dispersion relation which is non-monotonic in momenta produces
  what appears to be non-thermal behavior.
\end{abstract}


\end{frontmatter}

\section{Introduction}
\label{sec:intro}

The phase diagram of QCD at nonzero chemical potential and temperature is one of the subjects
of fundamental importance for the collisions of heavy ions.
While at zero baryon density the chiral transition is only crossover,
in the chiral limit for two massless flavors the transition could be of second order.
As the baryon chemical potential increases, the transition could become first order.
This would imply the existence of a tricritical point, where the lines of second and first
order transitions meet.  Away from the chiral limit, the tricritical point becomes a critical
endpoint \cite{Asakawa:1989bq,Stephanov:1998dy}.

In this paper we consider a possibility that besides a critical endpoint, that the phase diagram
could also include a Lifshitz regime.  In such a regime, there are spatially inhomogeneous
condensates.  The basic point of this note, however, is that the region in which effects
of the Lifshitz regime can affect experimental signals is {\it much} larger than the region
in which these condensates occur.  We suggest the region in which effects
of the Lifshitz regime occur are much larger greater than those of a tricritical/critical endpoint,
and show how this leads to what is apparently non-thermal behavior.

\section{Effective Lagrangians}

The effective Lagrangian is a sum of two terms,
$  {\cal L} = {\cal L}_0 + {\cal L}_{\rm HD}$,
where ${\cal L}_0$ has the usual form,
\begin{equation}
  {\cal L}_0 = \frac{1}{2} \left( \partial_0 \vec{\phi}\right)^2 
  + \frac{Z}{2} \left( \partial_i \vec{\phi}\right)^2 - h \vec{\phi} \cdot \vec{\phi}_0 + \frac{1}{2} m^2 \vec{\phi}^{\, 2}
  + \frac{1}{4} \lambda (\vec{\phi}^{\, 2})^2 
  + \frac{1}{6} \kappa (\vec{\phi}^{\, 2})^3 \; .
  \label{lag0}
\end{equation}
The potential is standard.  In three spatial dimensions, which is relevant to phase transitions at nonzero temperature,
the six-point coupling $\kappa$ is dimensionless, and so renormalizable.

What is of interest are the kinetic terms.  In order to have a causal theory, there can only be terms quadratic
in the time derivatives.  For the spatial derivatives, however, we introduce the wave function renormalization
constant, $Z$.  In vacuum Lorentz (or Euclidean) invariance implies that $Z = 1$.  In a medium, however,
generally $Z \neq 1$.  What is unusual is that we will assume that $Z$ can be negative.

When $Z < 0$, it is necessary to add terms with higher derivatives to stabilize the theory
\begin{equation}
{\cal L}_{\rm HD} = \frac{1}{2 M^2} \left( \partial_i^2 \vec{\phi}\right)^2
+ \frac{1}{M_1} \vec{\phi}^{\, 2}  \left( \partial_i \vec{\phi}\right)^2 + \frac{1}{M_2} (\partial_i \vec{\phi}^{\, 2})^2 \; .
\label{laghd}
\end{equation}
These are all non-renormalizable couplings.  The dimensions are those for a scalar field in three dimensions, where
$\phi \sim \sqrt{{\rm mass}}$.  In four dimensions, where $\phi \sim {\rm mass}$,
the coefficients of these terms would be $1/M^2$, $1/M_1^2$, and $1/M_2^2$, respectively.

If $Z$ is positive, then the usual analysis goes through.  Start first with the chiral limit, $h=0$.  
When the quartic coupling is positive, $\lambda > 0$, then
there is a second order phase transition as $m^2$ vanishes.  The six-point coupling $\kappa$ is assumed to be positive,
and thus ensures stability.  When the quartic coupling $\lambda$ is negative, there is a first order transition when
$m^2$ is positive.

When $m^2$ and $\lambda$ both vanish, there is a tricritical point.  The critical fluctuations at this point are
controlled by $\kappa$, and runs logarithmically when $\lambda$ vanishes.  This generates logarithmic corrections
to the critical exponents of mean field theory.

Away from the chiral limit, $h \neq 0$, the second order transition is washed out, and for sufficiently small $h$,
becomes cross-over.  The line of first order transitions persists, although it does become weaker and shifts.  The point
is that this must then end in a critical endpoint.

As $\lambda$ can become negative, the same is true for $Z$.  This is not so surprising.  A fermion loop has
a negative sign relative to that for a boson loop, so that at large quark chemical potential,
this can dominate.  Indeed, that $Z < 0$ at nonzero density is generic in $1+1$ dimensions \cite{Basar:2009fg}.

When the mass squared is negative, the theory lowers its energy by generating a condensate, which is constant in space.
When $Z$ is negative, the theory lowers the energy by generating a condensate in $(\partial_i \vec{\phi})^2$.  This
implies that not only the internal symmetry, but also rotational symmetry, is spontaneously broken.  The analysis
of Goldstone bosons is then more involved than if only the internal symmetry is broken \cite{Hidaka:2015xza,Lee:2015bva}.
There is always a phonon, which arises from the longitudinal degrees of freedom.

\section{Size of Lifshitz/critical regime}

\begin{figure}[!t]
  \centering
  \includegraphics[width=0.5\textwidth]{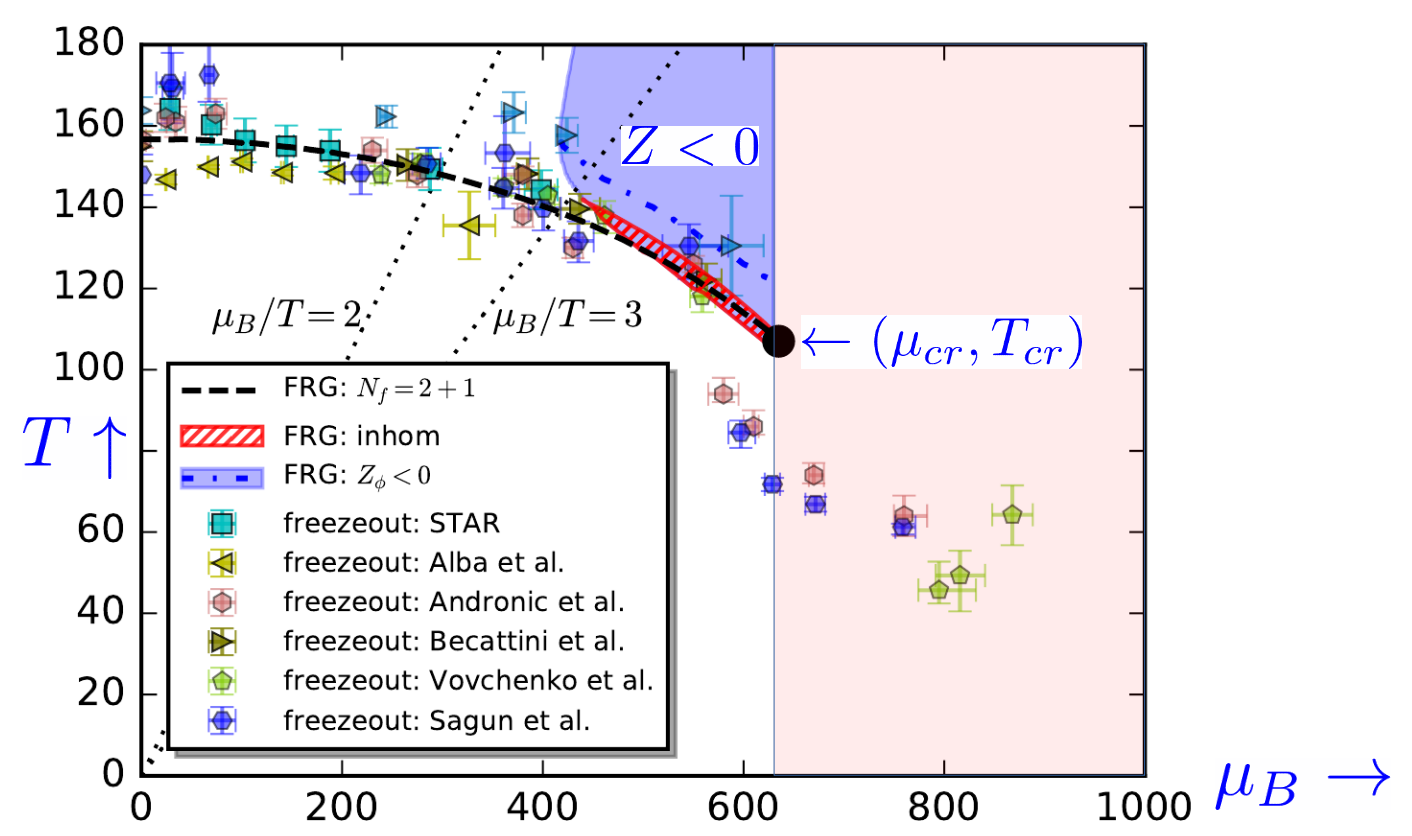}
  \caption{
    The phase diagram from the Functional Renormalization Group, Ref. \cite{Fu:2019hdw}.
    The critical endpoint is at $(\mu_{cr},T_{cr})$; the calculation only applies for $\mu < \mu_{cr}$ and
    $T> T_{cr}$.  The size of the black dot indicates the width of the critical
    region for the critical endpoint, and is very small.  In contrast, the (blue) shaded region denotes the
    Lifshitz regime, where $Z<0$, and is large.  The (red) hatched region is where the chiral
    condensate for light quarks is significant, and may be spatially inhomogeneous.
  }
\label{fig:FRG}
\end{figure}  
  
One might think that the width of the region for a critical endpoint could be arbitrary, and in particular,
rather wide \cite{Parotto:2018pwx}.
It is useful, though, to consider the chiral limit.  In this case, along the line of second order transitions the
width of the critical region is defined by the Ginzburg criterion.  It is approximately proportional to the
bare $\lambda$, and thus shrinks as the tricritical point is reached.  
In whatever direction in which one approaches the tricritical point, the width of the tricritical region is necessarily
small, vanishing as either $m^2$ or $\lambda$ vanishes.

In particular, for QCD the $\sigma$ meson is very heavy in vacuum.  Along the critical line, where $\lambda$ is
positive, the $\sigma$ becomes massless.  However, we are interested in the tricritical point, where $\lambda$ vanishes.
Thus the width of the tricritical region is defined by a Ginzburg criterion involving $\lambda$.  Without computing
in detail, it is necessarily narrow, overwhelmed by the critical region.

In the chiral limit pions are massless in the broken phase and at the critical point.  Away from the chiral limit
the pions are massive, and the only critical mode is the $\sigma$ at the critical endpoint.  
Again, the Ginzburg criterion can be computed, but presumably a necessary condition for the fluctuations
at the critical endpoint to become large is for the mass of the $\sigma$ to be small, at least near that
of the pion.  Even so, the width of the critical region for the critical endpoint is bound to be smaller than
for the tricritical point, and thus small.  This has been observed in mean field theory for some time \cite{Schaefer:2006ds}.

In contrast, consider the Lifshitz regime.  For a condensate to occur, it is not only necessary for $Z$ to be negative.
In addition, the dispersion relation for the pion must vanish at some nonzero momentum.  Further, a condensate is
not necessarily special to the chiral limit, and generically occurs as a type of Overhauser/Migdal
condensate \cite{Overhauser:1960,Brown:2007ara}.  What we emphasize later in this
note, however, is that physically measurable effects can occur as long as $Z$ is negative, even without the formation of
a condensate.

Since we need to compute the phase diagram at nonzero chemical potential, numerical simulations are of limited utility.
An analysis using the Functional Renormalization Group in QCD has been carried out
recently by Fu, Pawlowski,
and Rennecke \cite{Fu:2019hdw}.  The most interesting result is shown in Fig. \ref{fig:FRG}.
In accord with the general arguments above, the width of the critical region for the critical endpoint is {\it very} small.
This is because it is difficult the $\sigma$ is heavy in vacuum, and tends to remain so.

The region where $Z < 0$ is indicated by the (blue) shaded region.  Their analysis of $Z < 0$ is only valid for $T > T_{cr}$
and $\mu < \mu_{cr}$, where $(T_{cr},\mu_{cr})$ is the position of the critical endpoint.  In particular, it is {\it very}
difficult explicitly computing the nature of the phase with spatial inhomogeneities \cite{Buballa:2015awa}: it is necessary
to explicitly solve for a nonlinear, spatially varying field.  The region with such a chiral spiral
may occur even in the (red) hatched region, and certainly does for 
for $T < T_{cr}$ and $\mu \sim \mu_{cr}$.

What is most striking is that in a large region, the wave function renormalization $Z$ is negative.  
The analysis of Ref. \cite{Fu:2019hdw} does not apply for $T < T_{cr}$ or
$\mu > \mu_{cr}$, but it is very difficult to believe that,
simply by continuity, $Z<0$ does not also extend well into this region as well.

\section{Signals of the Lifshitz Regime}
\label{sec:lif}

\begin{figure}[!t]
  \centering
  \includegraphics[width=0.48\textwidth]{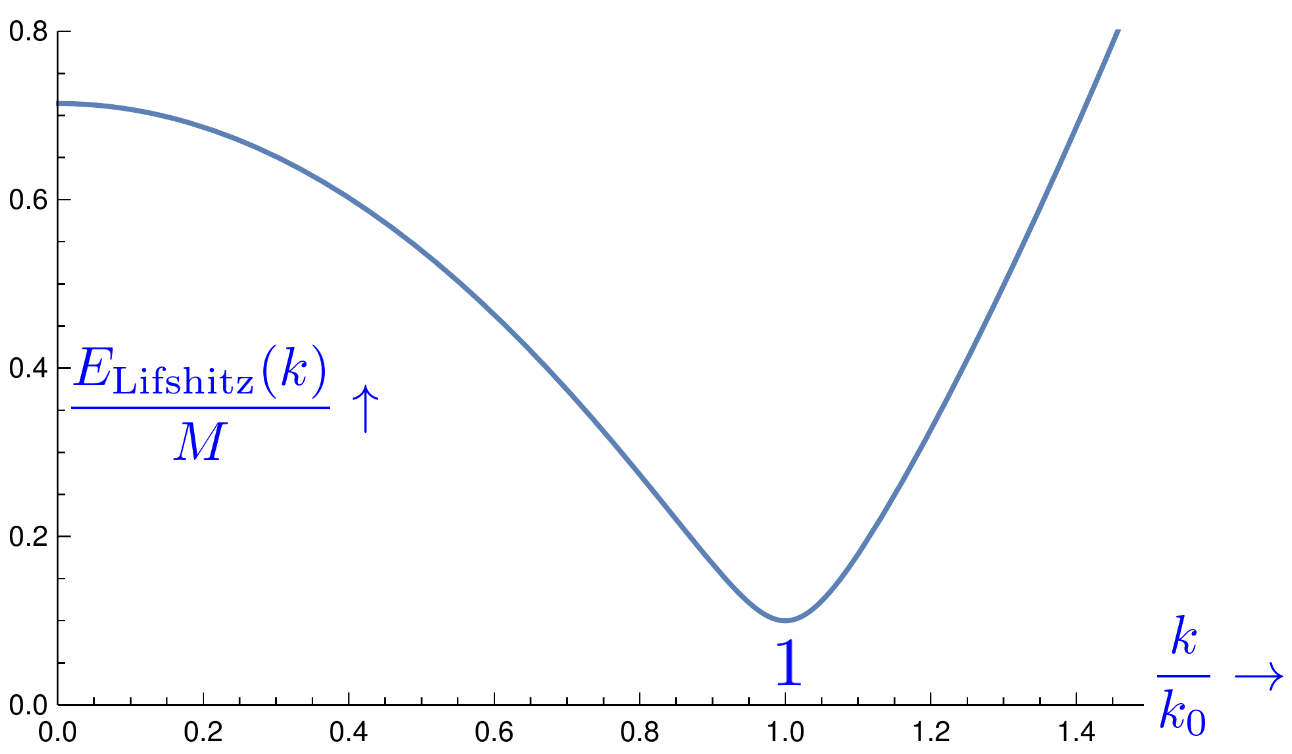}
  \includegraphics[width=0.48\textwidth]{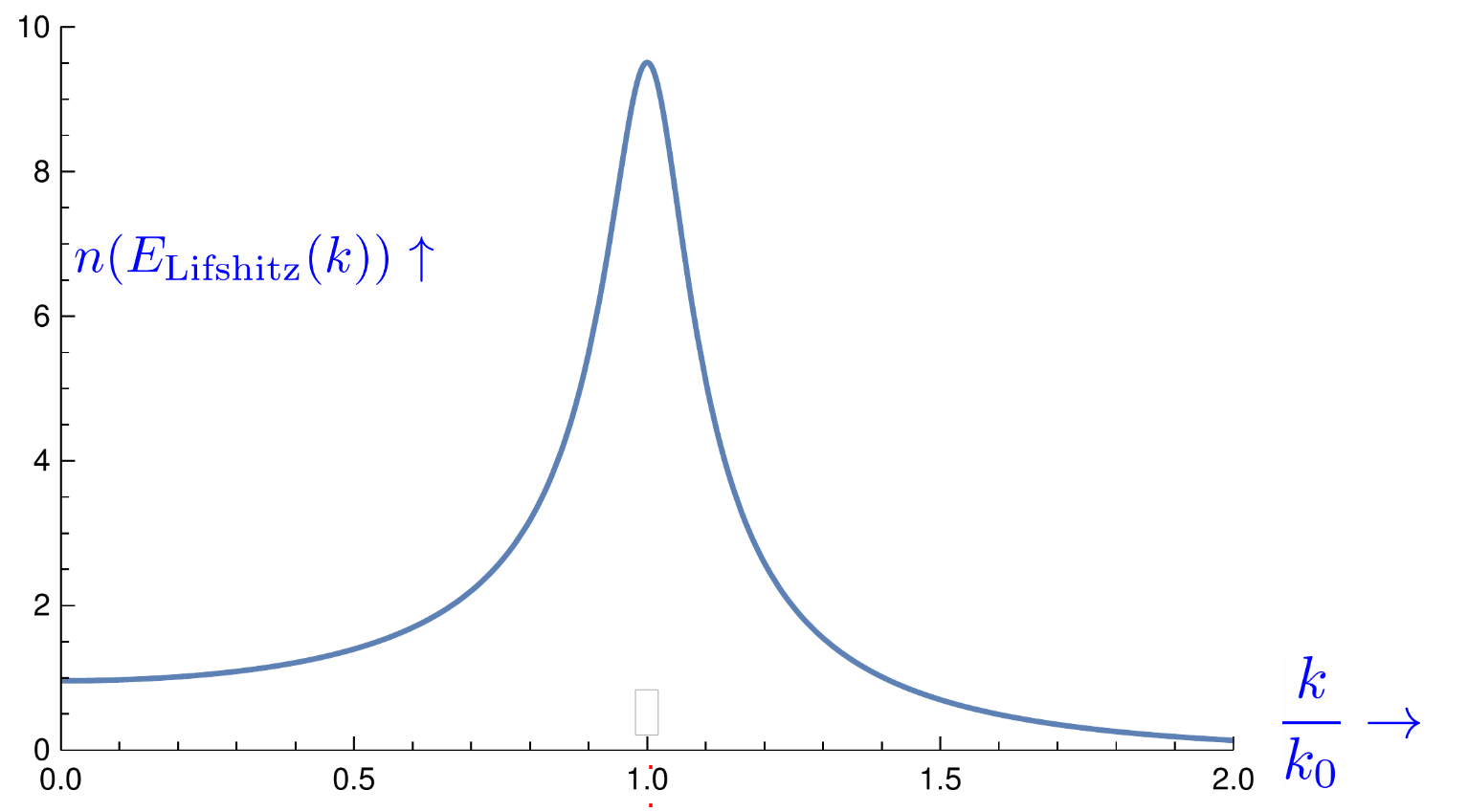}
  \caption{Left hand panel: the dispersion relation for the extreme example of Eq. \ref{test_disp}.  Right
    hand panel: the Bose-Einstein statistical distribution for this dispersion relation.}
\label{fig:lif_disp}
\end{figure}

How can one observe the Lifshitz regime, where $Z< 0?$  It is first necessary to minimize the effective
Lagrangians of Eqs. (\ref{lag0}) and (\ref{laghd}).  The simplest solution is a chiral spiral, characterized by
a momentum along a given direction, $k_0 \hat{z}$.  The fluctuations about this solution are unexpected,
and show that the Lifshitz regime is a type of ``quantum chiral liquid'' \cite{rdp}.  This analysis shows that
the dispersion relation for the fluctuations is given by

\begin{equation}
  \Delta_{\rm Lifshitz}(\omega,\vec{k}) = \frac{1}{\omega^2 + (\vec{k}^{\, 2})^2/M^2 + Z \vec{k}^{\, 2} + m_{\rm eff}^2} \; .
  \label{prop}
\end{equation}
To illustrate the effects possible, we take what is admittedly an {\it extreme} example, and consider
a field where the dispersion relation is
\begin{equation}
  E_{\rm Lifshitz}(\vec{k}) = \frac{(\vec{k}^2- k_0^2)^{\, 2}}{2 M^2} + m^2  \; ; \;
  M = k_0 = 10 \, m \; .
  \label{test_disp}
\end{equation}
This dispersion relation is illustrated in the left hand panel of Fig. \ref{fig:lif_disp}.  We
stress that this is {\it not} typical: by taking $M = k_0 = 10 m$, we are in a limit
close to where a spatially inhomogeneous condensate first arises.  Nevertheless, such an
extreme example is useful to demonstrate the type of effects which are possible.

Since this dispersion relation is positive everywhere, it does not lead to a spatially inhomogeneous condensate.
This would happen if there were a momentum where the energy vanishes.  In this case, though, $E(k_0) > 0$, and
it is simply a non-monotonic dispersion relation.

In Eucldean space-time, such as with numerical simulations on the lattice, it would direct to see this
non-monotonicity.  Typically, correlation functions fall off exponentially with distance, because the poles
of the propagator are purely imaginary.  With a propagator such
as Eq. (\ref{prop}), though, even purely spatial correlation function has poles which have both real and
imaginary parts.  This produces correlation functions which are are oscillatory function times an exponential,
which should be evident.

Similarly, a modified dispersion relation produce sharp differences experimentally.  For the dispersion
relation, the usual Bose-Einstein statistical distribution function is peaked not at zero momentum, but at
{\it non}zero momentum, about $k_0$: see the right hand panel in Fig. (\ref{fig:lif_disp}).  Needless to
say, since our example is so extreme, the differences need not be so dramatic.  Nevertheless, even small
deviations from the dominant success of thermal models
\cite{Andronic:2017pug} might be a hint to the presence of a Lifshitz regime.

The authors are funded by the U.S. Department of Energy
under contract DE-SC0012704.

\bibliographystyle{elsarticle-num}
\bibliography{lifshitz}

\end{document}